# Short-Term Prediction of Signal Cycle in Actuated-Controlled Corridor Using Sparse Time Series Models

**Bahman Moghimi,** Ph.D. candidate
Department of Civil Engineering, The City College of New York, New York, NY
E-mail: smoghim000@citymail.cuny.edu

**Abolfazl Safikhani,** Ph.D., Assistant Professor
Department of Statistics, Columbia University, New York, NY
E-mail: as5012@columbia.edu

**Camille Kamga,** Ph.D., Associate Professor
Department of Civil Engineering, The City College of New York, New York, NY
Email: ckamga@ccny.cuny.edu

**Wei Hao,** Ph.D.
Research Fellow, University Transportation Research Center,
The City College of New York, New York, NY
Email: hao@utrc2.org

**JiaQi Ma,** Ph.D. Assistant Professor
College of Engineering and Applied Science, University of Cincinnati, Cincinnati, Ohio
Email: jiaqi.ma@uc.edu

*Abstract:*

Traffic signals as part of intelligent transportation systems can play a significant role toward making cities smart. Conventionally, most traffic lights are designed with fixed-time control, which induces a lot of slack time (unused green time). Actuated traffic lights control traffic flow in real time and are more responsive to the variation of traffic demands. For an isolated signal, a family of time series models such as autoregressive integrated moving average (ARIMA) models can be beneficial for predicting the next cycle length. However, when there are multiple signals placed along a corridor with different spacing and configurations, the cycle length variation of such signals is not just related to each signal's values, but it is also affected by the platoon of vehicles coming from neighboring intersections. In this paper, a multivariate time series model is developed to analyze the behavior of signal cycle lengths of multiple intersections placed along a corridor in a fully actuated setup. Five signalized intersections have been modeled along a corridor, with different spacing among them, together with multiple levels of traffic demand. To tackle the high-dimensional nature of the problem, penalized least squares method are utilized in the estimation procedure to output sparse models. Two proposed sparse time series methods captured the signal data reasonably well, and outperformed the conventional vector autoregressive (VAR) model - in some cases up to 17% - as well as being more powerful than univariate models such as ARIMA.

## Introduction and Literature Review

Traffic signals are one of the most significant components of the emerging system of smart cities. They have been designed to control the demand in a way to improve traffic flow and reduce crashes in urban networks. There have been many developments of controlling logics (List & Mashayekhi, 2016). The first generation of signal control logics was Pre-Times (Fixed) signal control. In fixed control systems, the value of the green time and cycle length were fixed regardless of demand variation, which could induce a lot of slack time - the green time that is not being used by system users. Although some attempts have been made to have different fixed-time signal logics by time of day (a.m. and p.m. peak, midday, night time, etc.), this approach still imposes a lot of slack time on users. Subsequently, adaptive signal control has been designed to use information from the historical data of the past 5 or 10 minutes to update signal control parameters (including cycle length, split, and offset) and optimize timing and phasing to reduce user delays. There have been many adaptive control packages developed (Bing & Carter, 1995; Luyanda et al., 2003; Brilon & Wietholt, 2013). At the same time, with the rise of intelligent technologies such as detectors, sensors, wireless communication, vehicle-to-vehicle, and vehicle-to-infrastructure communications, signals have recently been designed intelligently using actuated control (Cesme and Furth, 2011; Agbolosu-Amison et al., 2012). Actuated control was presented to capture the demand at every second or even at a tenth of a second.

Fully actuated logic controls traffic signals in real time. It captures the underlying characteristics of the demand every second or even every tenth of a second. The control system collects traffic data through sensors, loop detectors, video, or radar. The actuated signal matches supply to demand in real time. It has the feature of compensation which means if a phase assigns more time to one direction because of reasons like emergency evacuation or transit signal priority, it compensates and allocates more time to the conflicting phases in the next cycle. The efficient fully actuated control operates so quickly that it results in shorter splits, cycle lengths, and therefore less delay to users (Cesme and Furth, 2011). Features enabling actuated signal control to operate as quickly as possible include: 1) using upstream/extension detectors, rather than over stop-line detectors, for gap detection, 2) using non-simultaneous gap-out logic instead of simultaneous gap-out 3) having shorter critical gaps, and 4) having shorter minimum green times (Furth et al., 2010). Since, in actuated control, the signal cycle length is changing from time to time; this logic makes it difficult to provide good coordination among signals placed along a corridor.

Fully actuated control for an isolated intersection has demonstrated high flexibility and capability to include transit priority, and other real-time traffic management techniques such as emergency evacuation. In addition, having a more accurate prediction of the actuated signal cycle length can be useful when it comes to intelligent traffic management, predicting travel time, and transit signal priority (TSP). For instance, transit signal priority attempts to change/adjust signal timing in order to turn the signal green for transit vehicles, resulting in less delay to transit. It has been shown by many studies (Sun et al., 2007; Moghimidarzi et al., 2016;

Wadjas and Furth, 2003; Li et al., 2012) that having longer prediction of transit arrival times can provide better conditions for transit to move faster. Thus, with the knowledge of a longer horizon of transit arrival times and more accurate signal cycle predictions, actuated control logics can gradually, rather than abruptly, change signal phases as the transit vehicle approaches the target intersection. Such a predictive approach results in greater reduction in transit delay without disrupting general traffic, and improves transit's travel time, crowding, and reliability.

Although actuated signal control has a lot of benefits, the prediction of the next cycle length is not an easy task since it varies from cycle to cycle. It is cumbersome to predict actuated signal cycle for an isolated signal and even more difficult to predict when there are multiple signals placed along a corridor. With regard to estimating signal cycle length for actuated control, Lin (1982) developed a deterministic model to estimate the average green time and cycle length using headway distribution for estimating the extension green period. As a continuation of Lin's research, Akcelik (1994) presented a model to estimate average green time with respect to minimum green, queue clearance time, and green extension time. More recently, Furth et al. (2010) introduced a new model to estimate actuated signal cycle length based on lost times. Seven lost time components were introduced. The evaluation of cycle length behavior on an isolated signal is performed based on the change of different levels of demand, detector setback, critical headway, and number of lanes per approach. Wadjas and Furth (2003) used a simple adaptive logic that takes the average of the last five cycles to predict the next value of cycle length and thereby used the predicted value to adjust the signal cycle through compression or expansion to provide priority for transit.

In this research, a multivariate time series model is presented to predict the next values of cycle length for each signal located along a corridor. Time series models have been deployed in transportation research studies in the last decades. Univariate time series modeling has been used in the study by Barua et al. (2015) to predict the traffic arrival demand; and William and Hoel's study (2003) predicted seasonal variation of freeway traffic conditions. Meanwhile, multivariate time series modeling has been taken into consideration in transportation-related problems such as forecasting the relative velocity in roads (Kamriankis and Prastacos, 2003), and predicting the traffic speed on a downstream link (Duan et al., 2016). In this article, with the use of sparse multivariate time series modeling, the prediction of signal cycles is presented for multiple traffic lights in a corridor.

In the next section, we describe how fully actuated control is programmed and implemented in this study. Then, the structure of time series models from univariate ARIMA models to multivariate VAR models, and further, sparse time series models are introduced. Subsequently, the outputs from all of the time series models for different levels of demand and spacing are discussed and analyzed. Finally, the last section presents the main findings, conclusions, and discusses future work.

# Methodology

A fully actuated control logic for five signalized intersections in a corridor is developed in this research. The developed controlling logic is based on standard actuation which turns the phases from green to red if the detected gap is larger than the minimum critical gap (gap-out), or exceeds the maximum green (max-out). The developed actuated control uses an upstream detector and is based on non-simultaneous gap-out logic. For left-turn phases, the logic skips the phase if there is no left-turn call. The logic uses a dynamic minimum green which updates the car counts between upstream and stop-line detector (cars-in-the-trap). It increments car counts by every upstream-detection and decrements it by every stop-line-detection.

Each traffic signal consists of many phases, some of which run concurrently and others that are in conflict which make up the critical phase. Each critical phase includes the green time and change interval (amber and all red time). Meanwhile, the duration of each phase can be a combination of some variables like the number of cars stopped during a red light, the stochastic arrival demand and platoon of cars dispatched from upstream intersections, the spacing between two adjacent signals, and other related variables. Therefore, in actuated control, each phase's queue length, split time, and obviously, signal cycle all changes from time to time. This research aims to determine if there is any covariance between the current value of a cycle length and its previous cycle lengths, and, more importantly, if there is any covariance among the datasets for different adjacent signals. For example, the study will investigate whether there is a correlation between the current value of the cycle length of a target intersection and its closest upstream signal or its next closest upstream signal, and so on. If there is any correlation among the datasets from different neighboring intersections, how we can make real time prediction with lots of parameters. To do so, the fully actuated control logic is being modeled using microsimulation traffic software called VISSIM, and then the cycle length data is considered from the time series modeling perspective to capture variability and correlation among signals.

*Time Series Model*

Time series models belong to a family of statistical models designed for data sets which are indexed by time. Time series models have been applied in different areas of finance, water resources, climate change, transportation, etc. (Montanari et al., 2000; Contreras et al., 2003; Lippi et al., 2013; Chen et al., 2010). The main objective of time series models is to capture the behavior of the data over time and to decipher the dependence among such data points in order to predict the future. Dealing with one data set, a univariate time series model such as the Auto Regressive Integrated Moving Average (ARIMA) model is a powerful yet simple statistical tool. On the other hand, when there are multiple datasets, such data should be treated using multivariate time series models. In the following subsection, the main concept of ARIMA models is briefly described and then multivariate models including VAR and sparse VAR models are introduced.

*ARIMA Model*

Suppose one has the data set $X_1, X_2, \ldots, X_n$, which are observed through time; i.e. $X_1$ is the observation at the first time point, $X_2$ is the observation at the second time point, etc. The Auto Regressive Moving Average ($ARMA$) model assumes that the current value of a time series is a linear combination of past observations and a linear combination of noises in the past observations. More specifically, the time series $X_t$ is called $ARMA(p,q)$ if

$$X_t - \phi_1 X_{t-1} - \cdots - \phi_p X_{t-p} = Z_t + \theta_1 Z_{t-1} + \cdots + \theta_q Z_{t-q} \qquad (1)$$

where $Z_t$ is considered as white noise with mean 0 and variance $\sigma^2 (WN(0,\sigma^2))$, $\phi_1, \phi_2, \ldots, \phi_p$ are called $AR$ constants, and $\theta_1, \theta_2, \ldots, \theta_q$ are called $MA$ constants. In the above model, the current data point depends on the past $p$ observations through $\phi_i$'s and the past $q$ observation noises through $\theta_i$'s. It is referred to (Brockwell and Davis, 2006) for more details about the properties of ARMA model.

*VAR Model*

Suppose $s_1, s_2, \ldots, s_k$ are $k$ fixed locations in $\mathbb{R}^d$ at which the response variable $\{y_t = (y_t(s_1), y_t(s_2), \ldots, y_t(s_k)) \in \mathbb{R}^k\}_{t=1}^T$ has been observed over a period of time with length $T$. Then, $y_t$ is called a VAR model if

$$y_t = \nu + \Phi^{(1)} y_{t-1} + \cdots + \Phi^{(p)} y_{t-p} + u_t, \ t = 1,2,\ldots,T, \qquad (2)$$

where $\nu \in \mathbb{R}^k$ the intercept, $\Phi^{(i)} \in \mathbb{R}^{k*k}$ the i-th lag coefficient matrix, and $\{u_t \in \mathbb{R}^k\}_{t=1}^T$ is a mean zero k-dim white noise with covariance matrix $\Sigma_u$. There are $k(k\,p + 1)$ parameters to estimate, and if $k$ is large compared to T, we may need to reduce the size in our estimation procedure. The linear regression compact matrix form of above formulation can be written as follows:

$$Y = \Phi Z + U \qquad (3)$$

where

$Y = [y_1 \ldots y_t] \quad (k*T);$ $\qquad \Phi = [\Phi^{(1)} \ldots \Phi^{(p)}] \quad (k*k.p);$

$z_t = [y'_{t-1} \ldots y'_{t-p}]' \ (k\,p*1);$ $\qquad Z = [z_1 \ldots z_T] \quad (kp*T);$

$U = [u_1 \ldots z_T] \quad (K*T);$ $\qquad \Phi_i^{(l:p)} = [\Phi_i^{(l)} \ldots \Phi_i^{(p)}](1*k(p-l+1)) \qquad (4)$

In order to deal with the high-dimensionality of the model when $k \gg T$, penalized least squares method are developed for parameters estimation. More specifically,

$$\widehat{\Phi} = argmin_\Phi \left\{ \frac{1}{2} \|Y - \Phi Z\|_2^2 + \lambda \, \Omega_i(\Phi) \right\}, \qquad (5)$$

where $\lambda$ is the tuning parameter to be selected by a rolling scheme ($0 < T_1 < T_2 < T$) (Song and Bickel, 2011; Nicholson et al., 2014; Nicholson et al., 2017), and $\Omega$ is the penalty function on the parameters $\Phi$.

For this study the following two penalty functions were chosen:

1) *LASSO – Least Absolute Shrinkage and Selection Operator*

The elementwise $L_1$ penalty known as LASSO (Tibshirani, 1996).

$$\Omega(\Phi) = \sum_{i=1}^{p} \left\| \Phi^{(i)} \right\|_1 \tag{6}$$

2) *HGLASSO – Hierarchical Groupe LASSO*

This penalty function (Nicholson et al., 2014; Nicholson et al., 2017) penalize the higher lag coefficients in a grouped way.

$$\Omega(\Phi) = \sum_{i=1}^{k} \sum_{j=1}^{k} \sum_{l=1}^{p} \left\| \Phi_{ij}^{(l:p)} \right\|_2 \tag{7}$$

Solving optimization problems with the form of equation (5) have been studied extensively under the penalty terms introduced previously (See (Tibshirani, 1996) and (Safikhani et al., 2017)) and references therein). Due to the hierarchical structure of the group penalties, especially in HGLASSO, this study applies the proximal gradient method introduced in (Jenatton et al., 2011). Further, the convergence rate of the proximal gradient method has been improved in (Beck and Teboulle, 2009) by introducing the Fast Iterative Soft-Thresholding Algorithm (FISTA). It is worth noting that the optimization problem (5) can be split over the rows of $\widehat{\Phi}$, which here is denoted by $\widehat{\Phi}_i$ for the $i-th$ row, $i = 1, \ldots, k$. This makes it possible to scale the computation even for high values of $k$ by parallel computing methods. In FISTA, a sequence of matrix coefficients $\widehat{\Phi}_i[r]$, $r = 1, 2, \ldots$ is introduced iteratively through

$$\hat{\phi} = \widehat{\Phi}_i[r-1] + \frac{r-2}{r+1}\left(\widehat{\Phi}_i[r-1] - \widehat{\Phi}_i[r-2]\right)$$
$$\widehat{\Phi}_i[r] = Prox_{s\lambda\Omega}\left(\hat{\phi} - s\nabla f_i(\hat{\phi})\right), \tag{8}$$

with $f_i(\Phi_i) = \frac{1}{2}\|Y_i - Z_i\Phi_i\|_2^2$, $\nabla f_i(\Phi_i) = -Z_i'(Y_i - Z_i\Phi_i)$ the vector of derivatives of $f_i(\Phi_i)$, $Y_i$ and $Z_i$ are the $i-th$ row of $Y$ and $Z$, respectively. $s$ is the step size (here we choose s to be $1/\sigma_1(Z_i)^2$ where $\sigma_1(Z_i)$ is the largest singular value of $Z_i$), and

$$Prox_{s\lambda\Omega}(u) = argmin_v \left(\frac{1}{2}\|u - v\|^2 + s\lambda\Omega(v)\right). \tag{9}$$

The proximal $ft(a)$ has a closed form for both penalty functions (6) & (7) defined above (See for example algorithm 2 in (Nicholson et al., 2014)). As for the tuning parameter selection, the time points are divided into three parts (usually equally distanced) $0 < T_1 < T_2 < T$. The estimation procedure for fixed values of $\lambda$ will be applied for the first part, i.e. $t = 1, 2, \ldots, T_1$. Then, the mean squared prediction error (MSPE) for predicting one step ahead is calculated over all $k$ time series components on the time

interval $[T_1 + 1, T_2]$:

$$MSPE = \frac{1}{k(T_2-T_1)} \sum_{i=1}^{k} \sum_{t=T_1+1}^{T_2}(Y_i(t) - P_{T_1} Y_i(t))^2 \qquad (10)$$

where $P_{T_1} Y_i(t)$ is the best linear predictor of $Y_i(t)$ based on the first $T_1$ observations. Now,

$$\hat{\lambda} = argmin_\lambda MSPE(\lambda) \qquad (11)$$

The model performance then can be quantified by the MSPE on the last part of the data, which is on the time interval $[T_2 + 1, T]$.

**General Results**

In order to understand signal cycle behavior in a corridor, one should first take a look at such behavior at an isolated intersection. Thereby, the behavior of the signal cycle of an isolated signal is briefly discussed here. It has been found (Moghimi et al., 2017) that the behavior of a signal cycle can be captured by using the univariate time series model, the ARIMA model. Figure 1 demonstrates the autocorrelation function (ACF) over signal cycle data, acquired for different levels of demand which were loaded on an isolated signalized intersection with two critical phases. As seen, as the demand increases from 800 veh/hr to 1400 veh/hr, the correlation between two consecutive signal cycles becomes more pronounced. This indicates that the behavior of signal cycle under fully actuated control can be analyzed through time series models, specifically when the level of demand is medium or high. Moreover, given that there is a significant correlation among datasets, a more precise prediction will be needed. Meanwhile, when there are left-turn phases, for signals with more than two critical phases, applying the linear regression over skipping indicators and then using ARIMA model could result in a smaller prediction error. For more details, the reader is referred to (Moghimi et al., 2017).

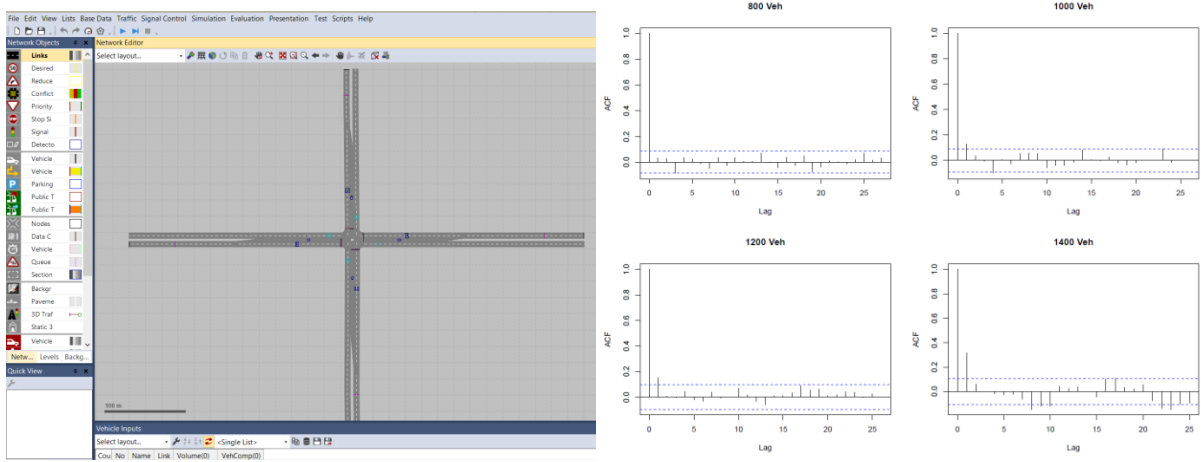

*Figure 1: Isolated actuated signal layout in VISSIM (figure on the left), and sample ACF over different levels of demand (figure on the right)*

In this paper, multivariate time series models on signal cycle data along a corridor are developed and analyzed. To do so, a corridor of five signalized intersections is modeled using

VISSIM microsimulation which is depicted in Figure 2. The signal control logic is programmed in C++, using VISSIM's application programming interface (API). In the developed model, at every time step of the simulation, detector information is passed from the simulation to the controller and then the signal phase state is returned to the simulation program. The corridor is an east bound one-way street with two lanes. Each controller runs under fully actuated control with two critical phases. To see the covariance of demand from one signal to another, the demand on cross streets is set as 600 veh/hr and the demand volume on west entry streets changes from 800 veh/hr to 1600 veh/hr, with the increment of 200 veh/hr, as a realization of how the street functions from off-peak to peak hour conditions. The minimum and maximum green times are defined as 12 seconds and 50 seconds, respectively. Upstream detectors were located about 2 seconds of travel time from the stop line. Various spacing is modeled in this research, including 200 meters, 500 meters, and 1000 meters. For each scenario, the VISSIM simulation was run for 5 hours following a 15-minute warm-up period. Subsequently, the time series models defined earlier were applied over the cycle length data coming from the traffic simulation.

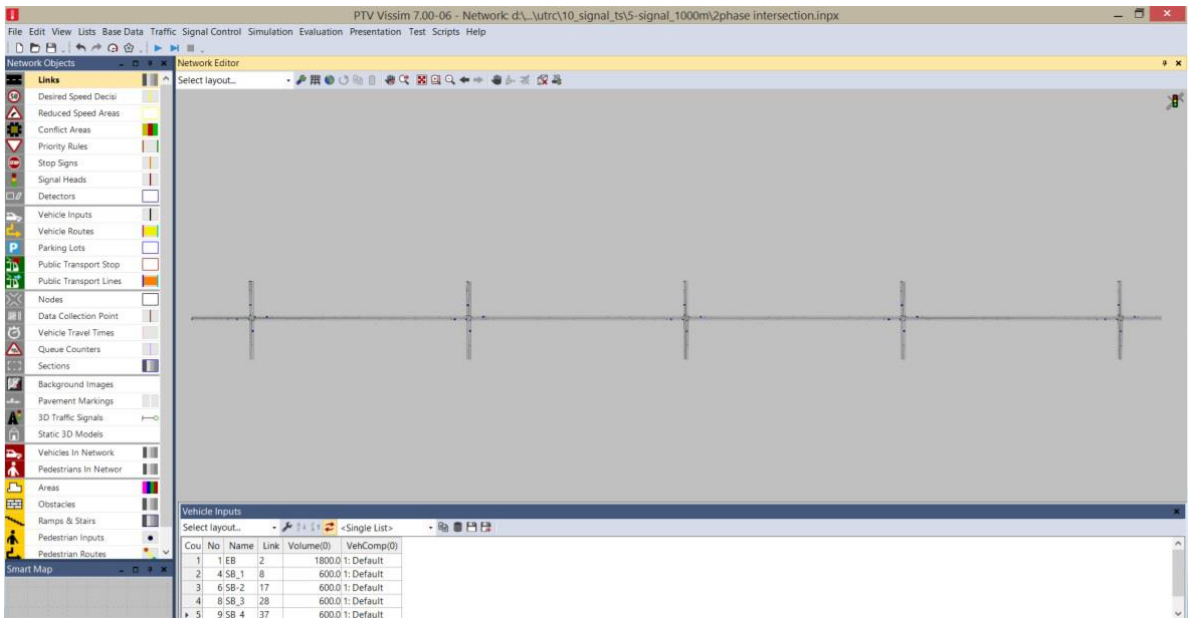

*Figure 2. Signalized corridor layout in VISSIM*

The prediction procedure is as follows: for each time series, the last 75 observations (the last 75 signal cycles) are excluded for the prediction purpose and then different prediction models are applied. Different prediction models are 1) averaging the last 5 cycles, 2) the univariate ARIMA model, 3) the multivariate VAR model, 4) the multivariate LASSO model, and 5) the HGLASSO model. To evaluate the performance of each method, a Mean Squared Prediction Error (MSPE) as shown in equation (10) is used. The lower the MSPE the better the model performance.

To evaluate the variation of signal cycles along the corridor, different scenarios were analyzed including a signalized corridor with spacing between signals varying from 200m, 500m, and 1000m, along with loading different levels of demand for each scenario. The prediction performance for the three spacing scenarios, 200m, 500m, and 1000m, are shown in Tables 1, 2, and 3, respectively. Noteworthy, the results shown in Table 1, 2, and 3 are the combined MSPE results of all 5 simulated intersections.

Table 1 shows the results of the five models with 200 meters spacing, and the levels of demand ranging from 800 veh/hr to 1600 veh/hr. As seen, all four time series models, at all levels of demand, perform better as compared to the conventional method of averaging the last 5 cycles. Moreover, the multivariate models, VAR, LASSO, and HGLASSO, outperformed the univariate model statistically; with the gaps between their values becoming wider as the demand increases. This is due to the fact that, as demand increases, the effect of neighboring intersections becomes stronger, and therefore the multivariate models VAR/LASSO/HGLASSO outperform the univariate model since they account for this effect.

*Table 1: MSPE of all models with 200 spacing between signals*

| Spacing | 200 meters | | | | |
|---|---|---|---|---|---|
| **EB Volume (v/hr):** | **800** | **1000** | **1200** | **1400** | **1600** |
| MSE.avg | 39.2 | 38.5 | 30.4 | 47.9 | 90.8 |
| MSE.univ | 31.4 | 32.0 | 25.2 | 40.4 | 76.3 |
| MSE.var | 26.9 | 31.7 | 22.4 | 36.8 | 66.9 |
| MSE.LASSO | 26.8 | 30.8 | 22.1 | 35.4 | 66.1 |
| MSE.HGLASSO | 26.8 | 30.8 | 22.1 | 35.4 | 66.1 |

*Table 2: MSPE of all models with different levels of demand having 500 meters spacing*

| | Spacing | 500 meters | | | | |
|---|---|---|---|---|---|---|
| | **EB Volume (v/hr):** | **800** | **1000** | **1200** | **1400** | **1600** |
| 1 lag | MSE.avg | 29.1 | 32.0 | 31.7 | 50.2 | 85.7 |
| | MSE.univ | 26.1 | 29.1 | 25.2 | 39.4 | 70.4 |
| | MSE.var | 26.0 | 28.2 | 17.3 | 31.0 | 58.0 |
| | MSE.LASSO | 25.2 | 27.5 | 17.3 | 31.0 | 56.5 |
| | MSE.HGLASSO | 25.2 | 27.5 | 17.3 | 31.0 | 56.5 |
| | | | | | | |
| 2 lags | MSE.univ | 26.5 | 29.1 | 25.6 | 39.6 | 70.4 |
| | MSE.var | 28.7 | 31.3 | 17.9 | 33.7 | 62.0 |
| | MSE.LASSO | 25.3 | 27.9 | 17.3 | 31.0 | 57.4 |
| | MSE.HGLASSO | 25.2 | 27.8 | 17.3 | 31.0 | 57.2 |

Table 2 reports the MSPE for the five models with 500 meters spacing at various levels of demand. As was found from simulation with 200m spacing, the time series models outperformed the averaging one; and the multivariate time series models (VAR and Sparse VAR models) have lower errors as compared to the univariate and averaging models. The LASSO and HGLASSO models have the smallest MSPE of all listed models.

*Table 3: MSPE of all models with 1000 meters spacing between signals*

|  | Spacing | 1000 meters | | | | |
|---|---|---|---|---|---|---|
|  | EB Volume (v/hr): | 800 | 1000 | 1200 | 1400 | 1600 |
| **1 lag** | MSE.avg | 29.7 | 23.6 | 33.7 | 53.2 | 69.8 |
|  | MSE.univ | 26.5 | 20.4 | 27.9 | 43.4 | 57.6 |
|  | MSE.var | 27.3 | 21.8 | 24.8 | 42.5 | 46.5 |
|  | MSE.LASSO | 26.0 | 20.2 | 24.5 | 41.9 | 46.7 |
|  | MSE.HGLASSO | 26.0 | 20.2 | 24.5 | 41.9 | 46.7 |
|  |  |  |  |  |  |  |
| **2 lags** | MSE.univ | 26.5 | 20.4 | 28.0 | 43.5 | 58.1 |
|  | MSE.var | 30.7 | 23.4 | 26.2 | 43.9 | 48.6 |
|  | MSE.LASSO | 26.4 | 20.3 | 25.3 | 40.7 | 45.5 |
|  | MSE.HGLASSO | 26.0 | 20.2 | 24.9 | 40.9 | 45.6 |
|  |  |  |  |  |  |  |
| **3 lags** | MSE.univ | 26.5 | 20.4 | 28.2 | 43.6 | 58.1 |
|  | MSE.var | 33.6 | 25.2 | 29.2 | 46.7 | 53.4 |
|  | MSE.LASSO | 26.6 | 20.4 | 25.4 | 40.5 | 46.0 |
|  | MSE.HGLASSO | 26.0 | 20.3 | 24.9 | 40.7 | 46.0 |

Table 3 reports the MSPE of the five models with a 1000m spacing between signals and various levels of demand. With a very large distance between signals, the platoon arrival pattern functions differently. In a sparsely-spaced signalized corridor, the platoon of cars is not as dense, large, and imminent as it is in a more closely spaced signalized corridor. For such cases, it is better to penalize some of the AR coefficients to achieve better prediction accuracy. Therefore, the Sparse VAR models can be more useful in this regard. It is noteworthy that, with 1000 m spacing, 2 or 3 time lags are better for the model since it takes 1 or 2 cycles on average for the platoon of cars to pass from the upstream intersection to the downstream one. The LASSO and HGLASSO models have similar performance when the lag $p = 1$ since there is no hierarchical effect in this case. As $p$ increases, the effect of the hierarchical penalty will be more apparent.

For example, for a spacing of 1000m, EB volume 800 and 1200, when $p = 3$, HGLASSO outperforms LASSO with the reduction of more than 2% in MSPE.

Considering the results shown in the tables above, it can be concluded that, when there is more than one signal, multivariate time series are a better tool for understanding the behavior of signal cycles. Meanwhile, it is found that the multivariate models used in this study, including VAR, LASSO, and HGLASSO, can achieve lower mean square prediction errors for signal cycle length as compared to the two other models - averaging the last five cycles and the univariate ARIMA model. Also, when the signals are closely spaced, such as 200 meters apart, the behavior of adjacent signals is highly correlated with one another; they are relatively behaving the same as compared to a corridor with signals spaced far apart. Meanwhile, the Sparse VAR models perform better in predicting the next value of signal cycle among the five models, when the distance between signals is high (1000 meters) and the demand is high as well. Such mathematical findings can be interpreted in practical research that has been conducted by transportation scholars on self-organizing signal control (Moghimidarzi et al. 2016, Furth and Cesme, 2013 & 2014). It indicates that when signals are closely spaced, those signals should be synchronized together by having the same cycle length. When they are spaced far apart, a platoon-based coordination can better provide coordination among signals.

**Particular Case**

*500 m spacing, EB volume is 1200 veh/hr and 1 lag*

Figure 3 shows the sample ACFs of the simulated signal cycle data for the 5 signals spaced 500 meters apart with an eastbound volume of 1200 veh/hr, which indicates the existence of a strong temporal dependency, especially between neighboring intersections. The graphs located along the diagonal represent the sample ACF for the data itself and the other graphs represent the correlation between two different signals. For instance, the graph labeled *"Srs 4 & Srs 5"* shows the correlation between the time series for signal 4 and that for signal 5. It is notable that the ACF on the signal 5 is more pronounced, which due to the fact that the simulation case is a one-way street and signal 5 is the last intersection receiving all traffic flows.

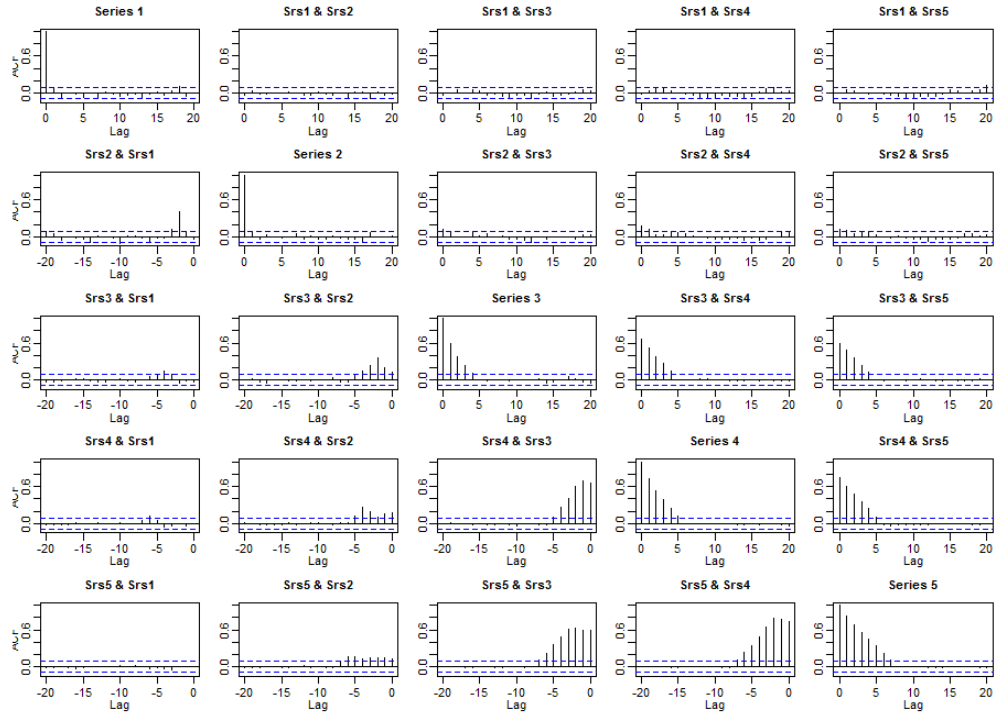

*Figure 3: Sample ACF of the five signals with 500 m spacing and 1200 veh/hr demand*

Considering the spacing of 500 meters and the speeds of the cars, it takes about a half cycle or one cycle for the cars to travel from one signal to the next. Hence, using the correlation of the previous two lags between signal cycles data can better predict the next value of the target cycle length. Figure 4 shows the $\Phi^{(i)}$ matrices among signals having two lags for both the VAR and HGLASSO models.

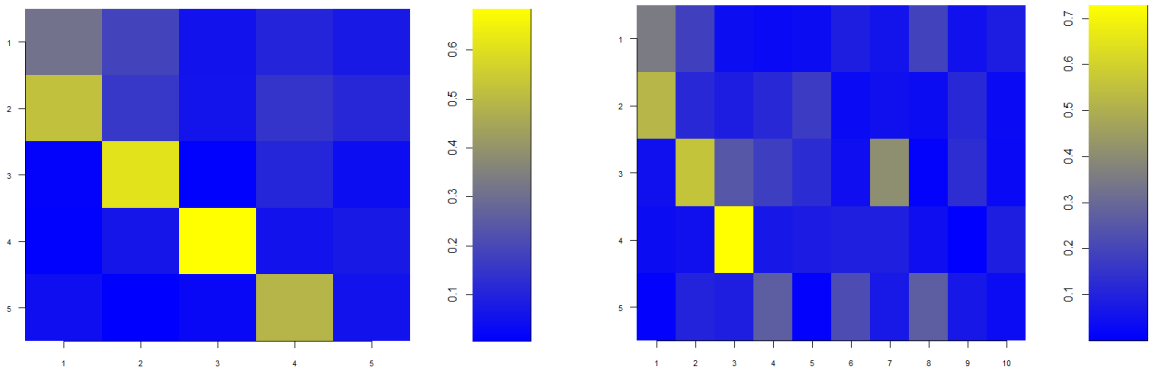

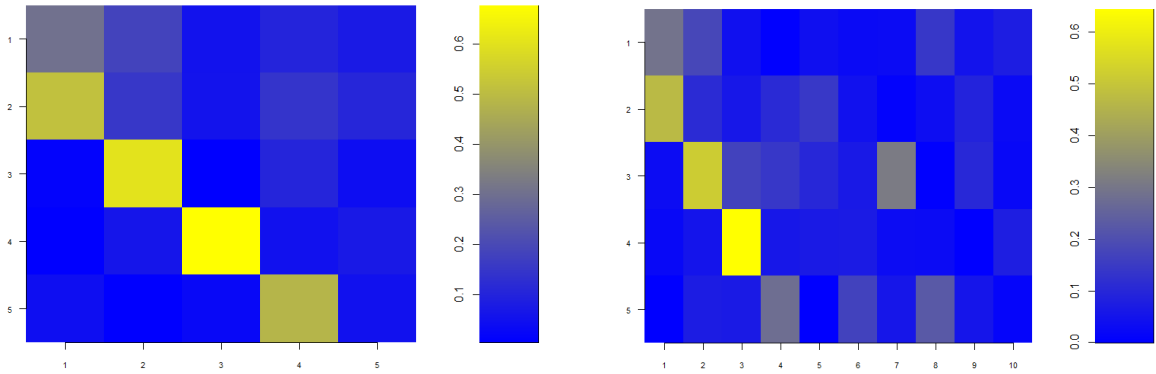

***Figure 4:*** *The top row shows the VAR coefficient matrix with one and two lags; and the bottom row shows the HGLASSO coefficient matrix with one and two lags*

Figure 5 shows the predicted values for the last 75 observations. The black line is the actual time series, the green line is the average of the last five cycles, the blue line is the *ARIMA* model, the pink line is the VAR model, the yellow line is the LASSO model, and the red line is the HGLASSO model. As can be seen, the multivariate models including the VAR, LASSO, and HGLASSO (pink, yellow, and red lines) perform well in capturing the true observations' oscillation.

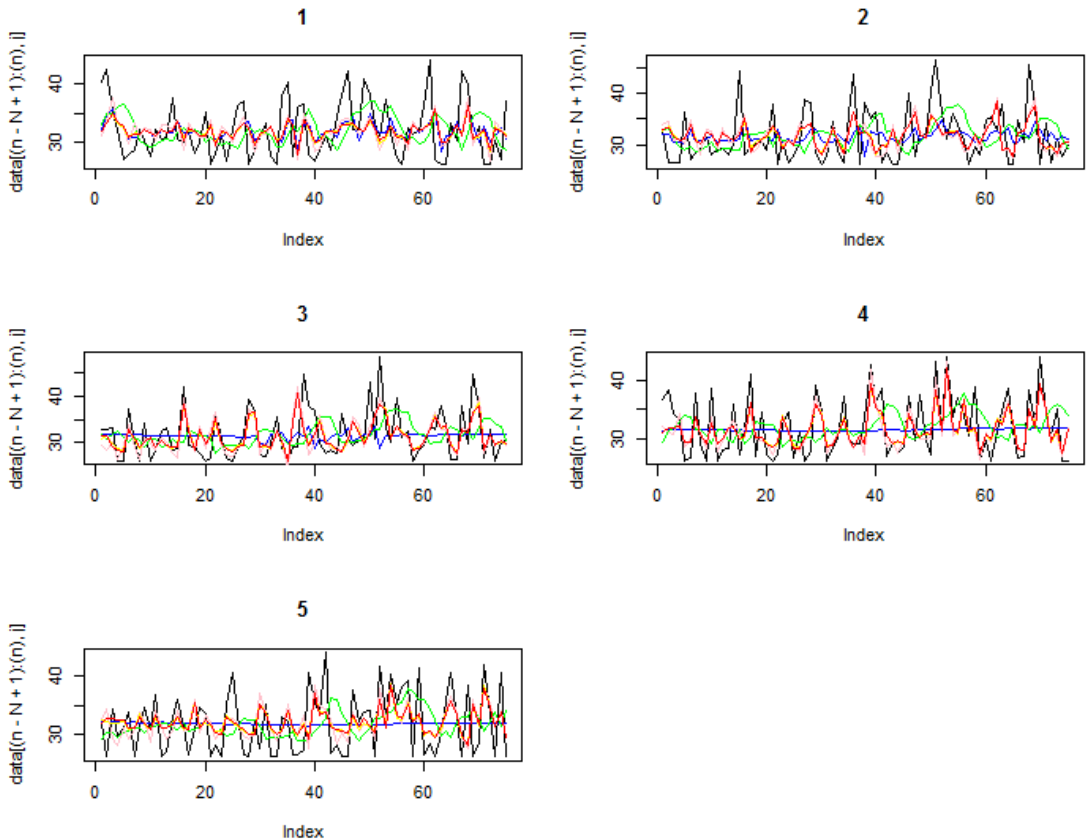

***Figure 5:*** *Predicted values of all prediction models for the five signals over the last 75 cycles*

*Results for Signal 5*

Since signal 5 is the last intersection in the one-way street simulation set-up shown in Figure 2, it has a more pronounced correlation from its upstream signals than the other signals do. Table 4 shows the mean squared prediction errors for the fifth signal. The results are for using one lag, two lags, and three lags in all of the prediction models for street-spacings of 200 m, 500 m, and 1000 m, respectively. As before, all the time series models perform better than the conventional cycle averaging model. Meanwhile, multivariate models outperform the univariate ARIMA model. Moreover, among the three multivariate models, the two sparse models in this paper - LASSO and HGLASSO – have lower prediction errors than the conventional VAR model. This reduction in MSPE is in some cases as high as 17% (the difference in the error from VAR model to LASSO model for the case of 1000 m spacing and 1600 v/hr volume)

***Table 4:*** *MSPE of all models for different levels of spacing and demand for the 5${}^{th}$ signal*

| Spacing | 200 m | | | | |
|---|---|---|---|---|---|
| EB Volume (v/hr) | 800 | 1000 | 1200 | 1400 | 1600 |
| MSE.avg | 49.3 | 53.7 | 34.5 | 38.1 | 76.5 |
| MSE.univ | 42.6 | 45.0 | 29.1 | 31.0 | 74.6 |
| MSE.var | 41.2 | 40.0 | 28.4 | 28.1 | 67.5 |
| MSE.LASSO | 39.0 | 38.9 | 27.6 | 27.5 | 65.6 |
| MSE.HGLASSO | 39.0 | 38.9 | 27.6 | 27.5 | 65.6 |

| Spacing | 500 m | | | | |
|---|---|---|---|---|---|
| EB Volume (v/hr) | 800 | 1000 | 1200 | 1400 | 1600 |
| MSE.avg | 35.6 | 37.8 | 28.4 | 86.7 | 84.0 |
| MSE.univ | 33.0 | 33.0 | 24.8 | 69.2 | 66.5 |
| MSE.var | 30.9 | 30.7 | 18.8 | 62.5 | 58.0 |
| MSE.LASSO | 30.1 | 32.7 | 17.6 | 57.5 | 51.9 |
| MSE.HGLASSO | 30.5 | 32.1 | 17.5 | 57.4 | 50.9 |

| Spacing | 1000 m | | | | |
|---|---|---|---|---|---|
| EB Volume (v/hr) | 800 | 1000 | 1200 | 1400 | 1600 |
| MSE.avg | 38.7 | 16.6 | 41.8 | 66.9 | 57.2 |
| MSE.univ | 32.4 | 14.9 | 37.4 | 58.0 | 50.4 |
| MSE.var | 38.7 | 15.4 | 39.8 | 55.9 | 45.5 |
| MSE.LASSO | 31.5 | 14.7 | 36.1 | 50.5 | 37.4 |
| MSE.HGLASSO | 31.1 | 14.8 | 35.2 | 49.7 | 39.6 |

**Conclusions and Future Studies**

This research analyzed the variation of signal cycles in a fully actuated set-up along a corridor for varying levels of traffic demand and signal spacing. The results derived from a VISSIM simulation for different scenarios were studied. It was found that time series models are strong statistical tools for capturing signal cycle behavior and predicting this behavior reasonably well, not only for isolated signals but also for multiple signals along a corridor. This research introduced a data-driven, statistical tool, the multivariate time series VAR, LASSO, and HGLASSO models, for predicting actuated cycle length for signals placed along a corridor. The results revealed that when there is a significant correlation between signal cycles, multivariate time series models can perform fairly well. Moreover, two sparse time series models, LASSO and HGLASSO, were found to outperform the conventional VAR model. The reason is that the sparse models penalized some of the statistical parameters and forced them to be zero to improve prediction accuracy. In the continuation of this research, a spatial-temporal autoregressive (STAR) method is going to be compared with the introduced sparse time series models. Furthermore, the use of accurate cycle length prediction in actuated signal control will be applied to predictive transit signal priority.

**Acknowledgment:**

This work is supported in part by NSF Grant IIS-1302423.